\documentclass{article}
\usepackage{amsmath}% if you are using this package,
\usepackage{amsthm}
\usepackage{graphicx}
\usepackage{latexsym}
\usepackage{amsfonts}
\usepackage{amssymb}

\begin{document}

\centerline{\bf Interpretation of Pulsed-Field-Gradient NMR in Terms of Molecular Displacements}

\begin{center}
George D. J. Phillies\\
Department of Physics\\
Worcester Polytechnic Institute\\
Worcester, MA 01609\\
phillies@wpi.edu\\
\end{center}

\centerline{\bf Abstract}

The relationship for Pulsed-Field-Gradient NMR between the amplitude $I(t)$ of the spin echo and the molecular displacement ${\bf X}(t)$ is examined.  $I(t)$ of a single species in a simple solution is determined by the mean-square displacement $\overline{X(t)^{2}}$.  With polydisperse species, or molecular probes in complex fluids showing memory effects, $I(t)$ in general includes large contributions from all higher even moments $\overline{X(t)^{2n}}$.  Conditions under which the NMR signal is indeed determined by the molecular mean-square displacement are noted.  A diagnostic that sometimes identifies when these conditions are not met is presented.

\section{Introduction}

Pulsed-Field-Gradient Nuclear Magnetic Resonance has proven to be an effective means of studying diffusion in simple and complex liquids. The quantity directly measured is the dependence of the amplitude of the spin echo on the strength and time separation of the gradient pulses.  In simple non-viscous liquids, the strength of the spin echo corresponds, on a molecular level, to the mean-square molecular displacement and to the molecular self-diffusion coefficient. In complex fluids, matters change.  The objective of this paper is to treat these changes on a fundamental level, obtaining relationships between spin echoes and the full probability distribution for single-molecule displacements.

The starting point is the relationship between the intensity $I(t)$ of the spin echo, the experimental parameters, and the molecular properties\cite{stejskal1965}.  The experiment applies short-duration magnetic field gradients ${\bf G}$  separated by a time interval $\Delta$, gradients enduring for a duration $\delta$.  The significant time $t$ is related to $\Delta$ and $\delta$ by $t = \Delta - \delta/3$.  The distribution function for molecular single-particle displacements ${\bf X}$ during time $t$ is $P({\bf X}, t)$.   The relaxation of $I(t)$ with increasing $t$ is related to molecular motions by\cite{walderhaug2010}
\begin{equation}
       I(t)/I(0)  = \int  d{\bf X}  P({\bf X}, t) \exp( 2 \pi \imath {\bf q} \cdot {\bf X}),
       \label{eq:relaxation1}
\end{equation}
the wave vector being
\begin{equation}
       {\bf q} =  {\bf G} \gamma \delta/ (2 \pi).
       \label{eq:qdefinition}
\end{equation}
where molecules of interest have gyromagnetic ratio $\gamma$.

Callaghan and Coy\cite{callaghan1992} provide a slightly different but equivalent expression for the time dependence of $I(t)$, which may be written
\begin{equation}
      g^{(1s)}(q,t) = N^{-1} \left\langle \sum_{n=1}^{N} \exp[ 2 \pi \imath {\bf q} \cdot {\bf X}_{n}(t) ] \right \rangle,
      \label{eq:relaxation2}
\end{equation}
in which $g^{(1s)}(q,t)$ is the {\em self} part of the dynamic structure factor, and in which ${\bf X}_{n}(t)$ is the displacement of the $n^{\rm th}$ of the $N$ particles during the time interval $t$.

$I(t)$ and $g^{(1s)}(q,t)$ are representations of the incoherent (single-particle) dynamic structure factor.  In somewhat different systems with different time and distance scales, at least four experimental methodologies are in use to determine $g^{(1s)}(q,t)$.  In addition to magnetic resonance methods, $g^{(1s)}$ is obtained using (i) inelastic neutron scattering from dilute isotopically-tagged tracer polymers in a polymeric fluid\cite{higgins1994}, (ii) optical tracer diffusion in which light scattering spectroscopy is applied to dilute solutions of an intensely-scattering tracer polymer, e.g., polystyrene, in solutions of an isorefractive but possibly non-dilute matrix polymer, e.g. polyvinylmethylether:orthofluorotoluene\cite{lodge1989}, and (iii) optical probe diffusion in which light scattering spectroscopy is used to observe intensely-scattering probe particles such as polystyrene spheres moving through solutions of weakly-scattering probe polymers\cite{phillies1999}.  While these methods have very different physical bases, they all measure the same underlying correlation function, seen on the right-hand-sides of eqns.\ \ref{eq:relaxation1} and \ref{eq:relaxation2}. Theoretical considerations\cite{phillies2005} applied to interpret relaxation in any of these methods  therefore remain applicable for the other methods. (Note, however, that in treatments of neutron and light scattering the factors of $2 \pi$ in eqns.\ \ref{eq:qdefinition} and \ref{eq:relaxation2} are cancelled against each other; the {\em wave vector} of eqn. \ref{eq:qdefinition} would in light scattering be termed the {\em inverse wave length}.)

\section{Results}

We now consider relationships between $g^{(1s)}(q,t)$ and $P({\bf X},t)$ for progressively more complicated systems.  Conditions under which $g^{(1s)}(q,t)$ is determined by the mean-square displacement are identified.  A diagnostic that sometimes identifies when $g^{(1s)}(q,t)$ is not being determined by the mean-square displacement is presented.

The simplest ideal system involves a solution of identical dilute particles whose motions are described by the Langevin equation.  For this ideal case, the distribution of particle displacements has a Gaussian form
\begin{equation}
      P({\bf X}, t) = \left( \frac{2 \pi}{3} \left\langle ({\bf X})^{2} \right\rangle \right)^{-3/2} \exp( - 3 ({\bf X})^{2}/ 2\langle ({\bf X})^{2}\rangle),
      \label{eq:distributionbrown}
\end{equation}
with the brackets $\langle \cdots \rangle$ indicating an ensemble average. The mean-square particle displacements satisfy $\langle ({\bf X})^{2}\rangle = 6 D t$.

For identical Langevin particles, the displacements are also a Markoff process, meaning that the particle displacements during non-overlapping time intervals are independent from each other.  The Gaussian and Markoff properties are independent of each other:  A statistical process may be Gaussian but not Markoffian or may be Markoffian but not Gaussian.  For the ideal system,
Berne and Pecora\cite{berne1976} evaluate eq.\ \ref{eq:relaxation1} and compute $g^{(1s)}(q,t)$, finding
\begin{equation}
     g^{(1s)}(q,t) = \exp (- 4 \pi^{2} q^{2} \langle ({\bf X})^{2}\rangle),
     \label{eq:g1qtmeansquare}
\end{equation}
The relaxation of the incoherent structure factor is determined by the mean-square particle displacements.

\begin{figure}
\includegraphics{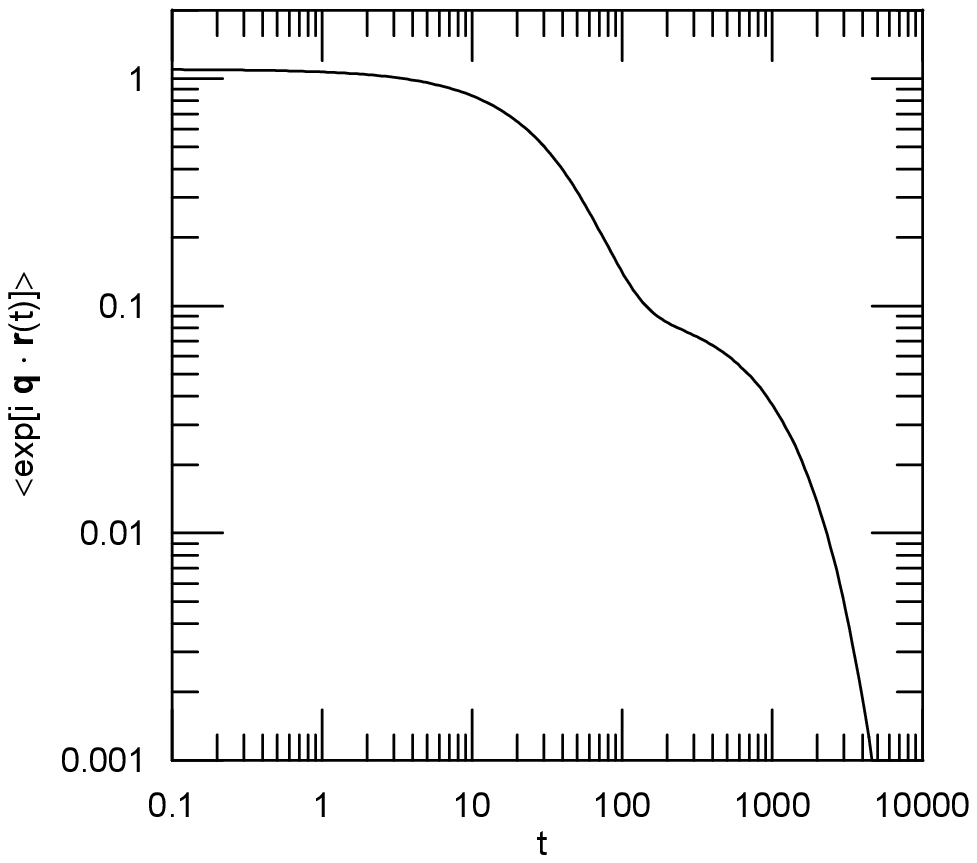}
\caption{Computed bimodal relaxation spectrum for a nominal mixture of two dilute Brownian species whose diffusion coefficients differ 30-fold.}
\label{figureone}
\end{figure}

\begin{figure}
\includegraphics{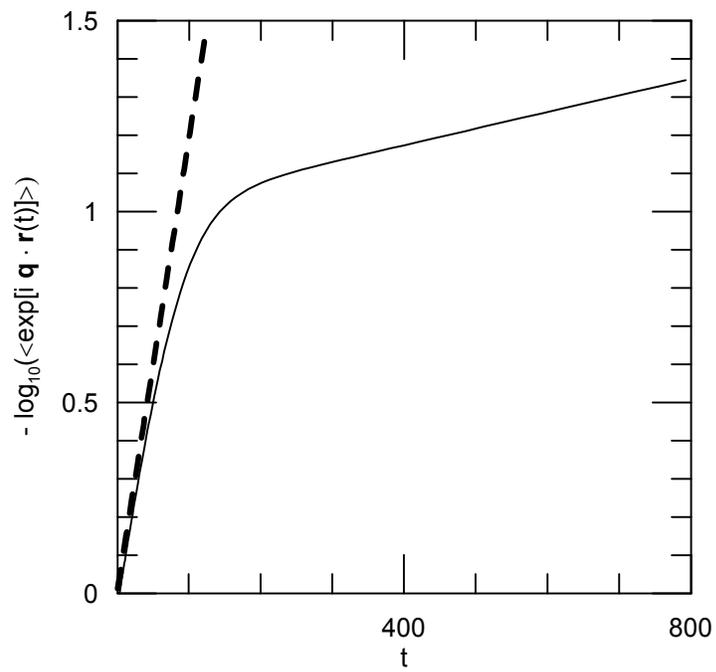}
\caption{Comparison of mean-square molecular displacement $\overline{X(t)^{2}}$ with PFGNMR spin-echo amplitude plotted as $- \ln(I(t))$ for the same mixture.}
\label{figuretwo}
\end{figure}

The extremely restrictive nature of the conditions under which eq.\ \ref{eq:g1qtmeansquare} is valid is often not emphasized. Figures \ref{figureone} and \ref{figuretwo} demonstrate the complete failure of eq.\ \ref{eq:g1qtmeansquare} when the conditions have been very slightly relaxed.  Figure \ref{figureone} shows $g^{(1s)}(q,t)$ computed for a nominal bidisperse system of Brownian particles.  Figure \ref{figuretwo}  plots (solid line) $- \ln(g^{(1s)}(q,t))/ (4 \pi^{2} q^{2})$, which would be the mean-square displacement if eq.\ \ref{eq:g1qtmeansquare} were correct.  Figure \ref{figuretwo} compares this form with the true mean-square displacement of the bidisperse particles (dashed line).  If eq.\ \ref{eq:g1qtmeansquare} were correct for this system, the two lines in the Figure would be exactly the same.  In fact, except that they share a common origin and initial slope, the lines are entirely different, because the conditions for the validity of eq.\ \ref{eq:g1qtmeansquare} are not satisfied by bidisperse Langevin particles. Particles of each species retain their sizes at all times, so ${\bf \Delta X}(t)$ is not a Markoff process.  Larger and smaller particles each persist in diffusing more rapidly or less rapidly, no matter for how long they are observed.  In terms of eq.\ \ref{eq:distributionbrown}, for each particle $P({\bf X}, t)$ is a Gaussian, but the average of $P({\bf X}, t)$ over all particles is not a Gaussian, but is instead the sum of two Gaussians.

The relationship between particle displacements, Gaussian processes, Markoff processes, and invocations of the Central Limit Theorem is made precise by Doob's theorem\cite{doob1942} and the discussion around it.  This matter has previously been analyzed in depth for light scattering spectroscopy\cite{phillies2005}, so key issues are only summarized here.  Note, however, that light scattering spectroscopy involves coherent scattering; correlations involving two different molecules at two different times contribute to the light scattering spectrum. Magnetic resonance involves the incoherent structure factor; terms in the light scattering treatment of $g^{(1)}(q,t)$ that involve two different molecules average in the magnetic resonance treatment exactly to zero. With this reminder, only a few key lines of the analysis of light scattering spectroscopy need to be repeated here.

Doob originally considered a random process $u(t)$, a random process being a time-dependent variable whose steps from time to time were random but had statistical properties.  $u(t)$ at a given time had a Gaussian random distribution, and its distribution at two times had a joint Gaussian distribution.  An example of $u(t)$ is the ${\bf q}^{\rm th}$ spatial Fourier component of the single-particle density
\begin{equation}
         a_{q}(t)  = \sum_{n} \exp(\imath 2 \pi {\bf q} \cdot {\bf R}_{n}(t))
         \label{eq:aqdef}
\end{equation}
Here ${\bf R}_{n}(t)$ is the location of particle $n$ at time $t$.  The remaining discussion is phrased in terms of $a_{q}$ rather than $u$.

Because the molecular locations are independently randomly distributed, all with the same distribution, and we are only considering the incoherent structure factor, the Central Limit Theorem guarantees that $a_{q}$ has a Gaussian random distribution.  For the incoherent structure factor, the Central Limit Theorem also shows that the displacements ${\bf X}_{n}(t) = {\bf R}_{n}(t) -  {\bf R}_{n}(0)$ are independent, all having the same distribution, so the incoherent part of $a_{q}(t)$ at any pair of times has a joint random distribution.

In the case that $a_{q}(t)$ is also a Markoff process, meaning that the probability distribution function for $a_{q}(t)$ at each of a series of times depends only on the value of $a_{q}(t)$ at the immediately prior time, Doob demonstrates that
\begin{equation}
      \langle a_{q}(0) a_{q}(t) \rangle  = \langle |a_{q}(0)|^{2} \rangle \exp (- \Gamma |t|),
      \label{eq:dooboutcome}
\end{equation}
except for the special case $\langle a_{q}(0) a_{q}(t) \rangle = 0$.  The demonstration applies equally to the coherent and incoherent parts of $\langle a_{q}(0) a_{q}(t) \rangle$.  For a Gaussian Markoff process, the relaxation is a simple exponential in $t$.

The Central Limit Theorem does not guarantee that a process has Markoff behavior, because the displacement of each particle during each time interval may be correlated with the displacement of the same particle during the prior time interval.  The absence of Markoff behavior is seen in velocimetry studies, in which the displacements of tracer particles during sequential time intervals are very heavily correlated and, correspondingly, $g^{(1s)}(q,t)$ shows not an exponential relaxation but instead, as it happens in this case, sinusoidal oscillations. Contrapositively, if the relaxation of $g^{(1s)}(q,t)$  is not a simple exponential, then the fluctuations in $a_{q}(t)$ are not described by a Gaussian Markoff process.  The lack of a simple exponential relaxation for $g^{(1s)}(q,t)$ is an infallible diagnostic that the assumptions leading to eq.\ \ref{eq:g1qtmeansquare} are not applicable in the system under consideration.

In systems in which $g^{(1s)}(q,t) = \exp(4 \pi^{2} q^{2} D t)$, there is no need to take short-time or small-$q$ limits, because the slopes of $\log(g^{(1s)}(q,t))$ as functions of $t$ or $q^{2}$ are the same at all $t$ and $q$.  Contrariwise, if experimentally $\log(g^{(1s)}(q,t))$ is not linear in $t$, Doob's theorem guarantees that the derivation leading to eq.\ \ref{eq:g1qtmeansquare} is not applicable to the system under study.

What happens in the non-ideal cases in which Doob's theorem is not applicable? Magnetic resonance and scattering both can measure $g^{(1s)}(q,t)$ as a function of $q$ or as a function of $t$.  In real applications, practical differences arise. Light scattering experiments observe light scattered through a single angle;  $g^{(1s)}(q,t)$ is automatically simultaneously computed at a large number of different times.  Observing $g^{(1s)}(q,t)$ at several $q$ with light scattering requires a disproportionate increase in experimental effort.  A single pulsed field gradient NMR cycle observes  $g^{(1s)}(q,t)$ for a single pair of values $q$, $t$. NMR studies can vary $t$, and can vary $q$ by changing $G$.  Leaving $t$ fixed avoids complications attendant to various relaxations\cite{walderhaug2010}, encouraging NMR studies to vary $q$ rather than $t$.

For small values of its exponential's arguments, and noting that symmetry eliminates terms odd in ${\bf X}$, a Taylor series expansion of the exponential in eq. \ref{eq:relaxation2} gives
\begin{displaymath}
          g^{(1s)}(q,t) =  1 - \frac{(2 \pi)^{2}}{2}  N^{-1}  \sum_{j=1}^{N} \left\langle ({\bf q} {\bf  X}_{j}(t))^{2}  \right\rangle
\end{displaymath}
\begin{equation}
              + \frac{(2 \pi)^{4}}{4!} N^{-1}  \sum_{j=1}^{N}  \left\langle ( {\bf q} \cdot {\bf  X}_{j}(t))^{4}  \right \rangle + {\cal O}(q^{6})
    \label{eq:g1qtexpand}
\end{equation}
the lead terms of which are
\begin{equation}
          g^{(1s)}(q,t) =  1 - \frac{(2 \pi)^{2}}{2} q^{2} \overline{X(t)^{2}}.
    \label{eq:g1qtlead}
\end{equation}
Here
\begin{equation}
        \overline{X(t)^{n}}  =    N^{-1}  \sum_{j=1}^{N} \left\langle ({\bf q} \cdot {\bf X}_{j}(t))^{n}  \right\rangle
   \label{eq:deltaXdef}
\end{equation}
Equation \ref{eq:g1qtlead} becomes accurate asymptotically in the limit of short time or in the limit of small wavevector.  The short time limit is approached by evaluating $\lim_{t \rightarrow 0} d  g^{(1)}(q,t)/dt$, as  by fitting $g^{(1)}(q,t)$ to a cumulant expansion\cite{koppel1972,frisken2001}.  The small-$q$ limit is approached by going to small angle or small field gradient.

If actually attained, the small-$t$ and small-$q$ limits are physically very different.  In the true small-time limit, particles have not had an opportunity to move very far; all relaxation process are still ongoing.  In the true small-$q$ limit, particles must displace through large distances to contribute to the relaxation, which takes a long time, so relaxation processes and memory-function effects will all have decayed.

It might be tempting to replace eqn.\ \ref{eq:g1qtlead} with a supposed exponential approximant
\begin{equation}
          g^{(1s)}(q,t) =  \exp\left[- \frac{(2 \pi)^{2}}{2} q^{2} \overline{(\Delta X(t))^{2}}\right].
    \label{eq:g1qtexponentiated}
\end{equation}
However, a full calculation shows that eqn.\ \ref{eq:g1qtexponentiated} is incorrect in its first non-trivial term, namely its Taylor series expansion is already wrong in its $q^{4}$ term. The error is revealed by noting that all quantities in $\exp(- \frac{(2 \pi)^{2}}{2} q^{2} \overline{X(t)^{2}})$ are constants that may be factored out of the ensemble average of eqn.\ \ref{eq:relaxation2}, yielding
\begin{displaymath}
       g^{(1s)}(q,t) =\left[ \exp(- \frac{(2 \pi)^{2}}{2} q^{2} \overline{X(t)^{2}})\right]
\end{displaymath}
\begin{equation}
      \times \left[ N^{-1} \sum_{j=1}^{N} \left\langle  \exp( 2 \pi \imath {\bf q} \cdot {\bf X}_{j}(t) + \frac{(2 \pi)^{2}}{2} q^{2} \overline{X(t)^{2}}) \right\rangle  \right].
       \label{eq:g1qtfirstexpansion}
\end{equation}

A Taylor series expansion of the second exponential leads to
\begin{displaymath}
       g^{(1s)}(q,t) = \left[\exp(- \frac{(2 \pi)^{2}}{2} q^{2} \overline{X(t)^{2}})\right]
\end{displaymath}
\begin{equation}
      \times \left[ (1 + q^{4}\frac{\overline{X(t)^{4}} - 3 ( \overline{X(t)^{2}})^{2}}{24} + {  \cal O}(q^{6})\right].
       \label{eq:g1qtsecondexpansion}
\end{equation}
The first correction appears at the $q^{4}$ level.  The exponential form is no more accurate that the linear form it was proposed to replace.

Iteration of the exponentiation process leads to a series
\begin{displaymath}
        g^{(1s)}(q,t) = \exp\left(- \frac{(2 \pi)^{2}}{2} q^{2} \overline{ X(t)^{2}}+ \frac{(2 \pi q)^{4}}{24} (\overline{X^{4}}-3 \overline{X^{2}}^{2})\right.
\end{displaymath}
\begin{equation}
        \left. - \frac{(2 \pi q)^{6}}{720} (\overline{X^{6}}-15 \overline{X^{4}} \ \overline{X^{2}}  +30 \overline{X^{2}}^{3}) +\ldots   \right).
       \label{eq:g1qtbigexpansion}
\end{equation}
Equation \ref{eq:g1qtbigexpansion} gives the relationship between the incoherent structure factor and various moments of the particle displacements.  The equation reduces to eqn.\ \ref{eq:g1qtexponentiated} if and only if $P({\bf X}, t)$ is a Gaussian, because in that case $\overline{X^{4}}-3 \overline{X^{2}}^{2}$ and corresponding terms of higher order in $X$ all vanish.  If $P({\bf X}, t)$ is a sum of Gaussians, as is the case for a polydisperse system, or is not Gaussian at all, eqn \ref{eq:g1qtbigexpansion} does not reduce to eqn.\ \ref{eq:g1qtexponentiated}.

Methods used to interpret light scattering spectra suggest two approaches to use equation \ref{eq:g1qtbigexpansion} to interpret intensity relaxations $I(t)$.

First, light scattering spectra, which are obtained at fixed $q$ and a large number of delay times, are routinely interpreted by means of time cumulants
\begin{equation}
      K_{n}   =   \lim_{t\rightarrow 0} (-q^{2})^{-n}  \frac{\partial^{n} \ln(g^{1s}(q,t))}{\partial t^{n}}
      \label{eq:timecumulants}
\end{equation}
$K_{1}$ is the light-scattering-intensity-weighted average diffusion coefficient $\overline{D}$, while $K_{2}$ gives the mean-square range of diffusion coefficients $\overline{D^{2}} - \overline{D}^{2}$. Similar identifications can be made for the higher time cumulants.  In practice, the $K_{n}$ are determined by fitting $g^{(1s)}(q,t)$ to a power series in $t$.  The number of cumulants that can be determined accurately is determined by the range of times studied and the signal-to-noise ratio, but is never large.

Second, by analogy with time cumulants, one may propose the use of wave-vector cumulants
\begin{equation}
      Q_{n}   =  (2n)! \left(\frac{-1}{2 \pi}\right)^{n}  \lim_{q\rightarrow 0} \frac{\partial^{n} \ln(g^{(1s)}(q,t))}{\partial (q^{2})^{n} }.
      \label{eq:Qcumulants}
\end{equation}
By direct calculation,
\begin{equation}
      Q_{1} =  \overline{X(t)^{2}}
      \label{eq:Q1def}
\end{equation}
is the mean-square displacement at time $t$, while
\begin{equation}
      Q_{2} =   (\overline{X^{4}}-3 \overline{X^{2}}^{2})
      \label{eq:Q2def}
\end{equation}
gives the deviation  from a Gaussian form of the displacement distribution.  If $P({\bf X}, t)$ has a simple Gaussian form, $Q_{2}$ and all higher wavve-vector cumulants $Q_{n}$, $n \geq 2$, vanish.  By analogy with the time cumulants, the second wavevector cumulant is usefully normalized as
\begin{equation}
     V = \frac{\sqrt{Q_{2}}}{Q_{1}}
\end{equation}
The third wavevector cumulant is
\begin{equation}
      Q_{3} =   (\overline{X^{6}}-15 \overline{X^{4}}\ \overline{X^{2}}  +30 \overline{X^{2}}^{3}).
      \label{eq:Q3def}
\end{equation}
Just as time cumulants are usefully obtained by fitting $g^{(1s)}(q,t)$ as a function of time to a power series in $t$, so also wavevector cumulants should usefully be accessible by fitting  $g^{(1)}(q,t)$ as a function of $q$ to a power series in $q^{2}$. Linear-least-square fits to power series will reliably determine the leading initial slopes of $g^{(1s)}(q,t)$ as a function of $t$ or $q^{2}$, even when the higher-order cumulants (and, hence, the degree of curvature of  $\ln(g^{(1s)}(q,t))$) are considerably different from zero.   Procedures for ascertaining the number of time cumulants that may actually be extracted from a given spectrum\cite{philliescumulants} should remain applicable as a path to determining how many wavevector cumulants can be extracted from a given $g^{(1s)}(q,t)$.

\section{Discussion}

In the above, the relationship between the pulsed-field-gradient spin echo intensity and the displacement distribution function of the observed molecules was discussed.  Only for monodisperse Brownian particles, for which  $g^{(1s)}(q,t)$ is a single exponential in $q^{2}$ and $t$, is $g^{(1s)}(q,t)$ determined by the mean-square particle displacement. If $g^{(1s)}(q,t)$ is not a single exponential in  $t$, $g^{(1s)}(q,t)$ receives contributions from the higher moments $\overline{X^{2n}}$, $n > 2$ of the displacement distribution.  In general $g^{(1s)}(q,t)$ depends on all higher even moments $\overline{X^{2n}}$ of the distribution function for particle displacements. An expansion for $g^{(1s)}(q,t)$ in terms of the moments is presented.  Analysis of $g^{(1s)}(q,t)$ in terms of time or wavevector cumulants was considered.

\pagebreak

\pagebreak

\end{document}